\documentclass[man]{apa7}
\usepackage[style = apa, uniquename=false,sortcites = true, sorting = nyt, natbib=true]{biblatex}
\usepackage{amsgen,amsmath,amstext,amsbsy,amsopn,amssymb}
\usepackage[utf8]{inputenc}
\title{Penalized Subgrouping of Heterogeneous Time Series}
\shorttitle{PENALIZED SUBGROUPING}

\author{
Christopher M. Crawford$^{1}$ \\
Jonathan J. Park$^{2}$ \\
Sy-Miin Chow$^{1}$ \\
Anja F. Ernst$^{3}$ \\
Vladas Pipiras$^{4}$ \\
Zachary F. Fisher$^{1}$ 
}
\affiliation{ 
$^{1}$ The Pennsylvania State University \\
$^{2}$ {University of California, Davis} \\
$^{3}$ University of Groningen \\
$^{4}$ The University of North Carolina at Chapel Hill 
}

\abstract{Interest in the study and analysis of dynamic processes in the social, behavioral, and health sciences has burgeoned in recent years due to the increased availability of intensive longitudinal data. However, how best to model and account for the persistent heterogeneity characterizing such processes remains an open question. The multi-VAR framework, a recent methodological development built on the vector autoregressive model, accommodates heterogeneous dynamics in multiple-subject time series through structured penalization. In the original multi-VAR proposal, individual-level transition matrices are decomposed into common and unique dynamics, allowing for generalizable and person-specific features. The current project extends this framework to allow additionally for the identification and penalized estimation of subgroup-specific dynamics; that is, patterns of dynamics that are shared across subsets of individuals. The performance of the proposed subgrouping extension is evaluated in the context of both a simulation study and empirical application, and results are compared to alternative methods for subgrouping multiple-subject, multivariate time series.}


\begin{document}
\maketitle

Recent technological advances have decreased the burden associated with collecting intensive longitudinal data (ILD) in the social, behavioral, and health sciences. The availability of such data has catalyzed interest in the study of constructs defined by the complex interplay between dynamic biopsychosocial processes. Despite these increases in both access to ILD and interest in the study of dynamic processes \parencite{Hamaker2017}, how best to model multivariate time series data arising from multiple individuals is still an open question. Central to this question is how researchers should best accommodate the persistent heterogeneity observed in many aspects of human behavior.  

Current methods for analyzing dynamic processes vary in the degree to which this heterogeneity is addressed \parencite[e.g.,][]{Liu2023}. Multilevel modeling approaches, for example, allow individuals to differ quantitatively on a limited set of dynamic features through the inclusion of random effects \parencite[e.g.,][]{Bringmann2013}. However, because standard multilevel models assume no qualitative differences in the pattern of relations among the dynamic processes, these approaches may be overly restrictive. For example, data generating processes characterized by individual differences in the patterning of zero and nonzero dynamics are poorly represented by such approaches. Moreover, violation of assumptions regarding the structure and distribution of random effects can adversely impact estimation and inference \parencite{McNeish2017}. Idiographic methods, conversely, allow for a great deal of flexibility through the specification of person-specific models (\cite{Lutkepohl2005}; \cite{Molenaar2004}). Indeed, recent work has shown that the dynamics of commonly studied constructs---such as depression and anxiety \parencite{Fisher2017}, externalizing behavior \parencite{Wright2015}, and personality \citep{Beck2020}---are characterized by marked heterogeneity in both the magnitude of the features and the patterning of dynamics, suggesting that person-specific methods may be needed to fully capture the complexity of the data generating processes. The increased flexibility afforded by idiographic approaches, however, may limit generalizability and inhibit the identification of shared dynamics critical for areas such as intervention development.

Recently, a number of approaches have emerged that offer alternatives for characterizing heterogeneity in dynamic processes by seeking to bridge the divide between nomothetic and idiographic approaches to modeling multiple-subject, multivariate time series. One of these approaches, the multi-VAR framework (\cite{Fisher2022}; \cite{Fisher2024}), is built upon the vector autoregressive (VAR) model and simultaneously estimates group- and individual-level models. Importantly, the multi-VAR approach accommodates both quantitative and qualitative heterogeneity in dynamics across individuals---that is, heterogeneity in both the magnitude and pattern of zero and nonzero dynamics---and is compatible with a number of penalization methods for structuring how information is shared across individuals. Currently, multi-VAR is only capable of estimating a single group-level model, presumably by all individuals. It may be the case, however, that for many processes shared patterns of dynamics also exist among subgroups, or clusters, of individuals. 

To address this limitation, the current project extends the multi-VAR framework to allow for data-driven identification of subgroups and penalized estimation of subgroup-level dynamics. The approach detailed herein is characterized by a number of advantageous features. First, both quantitative and qualitative heterogeneity are accommodated in the estimation of group-, subgroup-, and individual-level models. In contrast to most existing subgrouping frameworks, this allows for individual differences in the magnitude of dynamics both within and between subgroups. Moreover, these effects are not assumed to follow any specific distributional form, thereby capitalizing on the flexibility inherent in fully idiographic approaches while also prioritizing generalizability through the group- and subgroup-level models. Second, sparsity in the group-, subgroup-, and individual-level dynamics is induced through a penalized estimation procedure, addressing challenges associated with overparameterization in both single-subject \parencite[e.g.,][]{Sims1980} and multilevel approaches. Indeed, recent work on the development of a subgrouping approach incorporating finite mixture modeling into the multilevel VAR framework notes that estimation can be burdensome when specifying a large number of random effects \parencite{Ernst2024}. Finally, the simultaneous estimation procedure for group- and individual-level dynamics \parencite{Fisher2022} extends to subgroup-level effects. Conversely, iterative approaches to the identification and estimation of subgroup dynamics may encounter issues associated with similarly iterative variable selection methods, including overfitting and suboptimal solutions \parencite{McNeish2015}. These features are described in detail in the following sections.

The remainder of the article is organized as follows. First, we review existing methods for subgrouping multiple-subject, multivariate time series. We then introduce the multi-VAR framework and the penalized subgrouping extension. We conclude with a presentation of results from both a simulation study and empirical example conducted to examine the efficacy and utility of the multi-VAR subgrouping extension. 

\section{Subgrouping Methods for Time Series}

The identification of subgroups of individuals for whom shared patterns of dynamics exist, and the estimation of parameters that provide meaningful information about said subgroups, are of substantive interest in a variety of domains. Recent work, for example, has noted that current taxometric procedures for the diagnosis of psychopathological syndromes fail to account for the observed heterogeneity within putatively homogeneous diagnostic categories \parencite{Kotov2017}. Appropriately parsing such heterogeneity can accelerate the development of more effective treatment paradigms \parencite[e.g.,][]{Fisher2019}, echoing calls for more personalized approaches to diagnosis and treatment in both clinical science \parencite{Wright2020} and medicine \parencite{Hamburg2010}. Importantly, however, there is not a clear consensus regarding best practices for the identification of subgroups in multiple-subject, multivariate time series. Liao (\citeyear{Liao2005}), for example, notes that time series clustering methods can be organized according to which aspects of the data are used in the subgrouping procedure. For the purposes of the current work, we focus our attention on methods that derive clusters based on parameter estimates in the VAR framework.

One popular VAR-based clustering method is the alternating least squares (ALS) approach \parencite{Bulteel2016}, an iterative algorithm consisting of three steps. First, individuals are assigned to initial clusters, which can be done randomly or through the use of alternative clustering approaches, such as hierarchical clustering using Ward's criterion \parencite{Ward1963}. Next, a VAR model is fit to each cluster to obtain subgroup-specific parameter estimates. Finally, subgroup membership is updated by reassigning individuals to the cluster for which the sum of their squared prediction errors are minimized. These steps are repeated until there are no changes in subgroup membership. A notable limitation of this framework is that only between-cluster quantitative heterogeneity is modeled---that is, the ALS approach assumes \emph{a priori} that individuals within a cluster are governed by the same data generating process, and that there are no qualitative differences between clusters. Thus, different clusters are assumed to share the same pattern of zero and nonzero dynamics.

Work by Ernst and colleagues (\citeyear{Ernst2021}) relaxes the within-group homogeneity assumption with a VAR-based clustering approach that incorporates the Gaussian finite mixture model (GMM; \cite{Mclachlan1988}; \cite{Mclachlan2000}). Proceeding in a two-step fashion, person-specific VAR models are first fit to data for each individual, with the restriction that there are no between-person differences in lag order. A GMM is then applied to the person-specific parameter estimates, thereby allowing for quantitative differences in within-cluster estimates. With this added flexibility, however, comes increased complexity in both the estimation of the parameters that define the multivariate Gaussian distribution and selection of the model that best represents the structure of the data. Indeed, the use of relative fit statistics to compare models in the GMM framework often requires bootstrap-based methods \parencite{Mclachlan2000}. Simulation studies comparing these two VAR-based clustering methods have found that the GMM approach outperforms the ALS framework when within-group quantitative differences are present \parencite{Ernst2021}, whereas the opposite is true when the homogeneity assumption is met \parencite{Takano2021}.

Another approach is the subgrouping extension within the group iterative multiple model estimation (S-GIMME) algorithm \parencite{Gates2017}. The S-GIMME algorithm can be described in three sequential stages. In the group stage, the S-GIMME framework estimates a structural VAR model \parencite{Lutkepohl2005} for each individual. Modification indices \parencite{Sorbom1989} are used to identify paths to be added to the model if they significantly improve model fit for a majority of individuals, thereby returning estimates that are shared across individuals. Next, the subgroup stage uses these group-level results to construct a similarity matrix such that each element of said matrix represents the number of parameter estimates shared between two individuals in terms of sign and significance. The community detection algorithm Walktrap \parencite{Pons2006} is then applied to the similarity matrix to identify clusters of individuals with shared dynamics. Once subgroups have been identified, estimates for individuals within each subgroup are obtained in the same manner as in the group stage. Finally, the third stage estimates additional dynamics for each individual by iteratively adding paths until the model is deemed acceptable via commonly used fit indices and associated cutoff values \parencite{Lane2019}. The S-GIMME approach therefore returns estimates for each individual and whether each estimate is unique, subgroup-, or group-specific, thereby accommodating both quantitative and qualitative heterogeneity. Moreover, challenges associated with the estimation of VAR processes comprised of many variables are addressed through the use of stopping criteria at each stage, which promotes sparsity and parsimony. However, forward-selection procedures, such as S-GIMME, can be limited by their sequential nature, and thresholds for the fit indices used in the individual-stage search procedure have been shown to be inadequate in many settings \parencite{Mcneish2023}.

Finally, a recently developed subgrouping approach, the subgrouped chain graphical VAR (scGVAR), extends the graphical VAR framework \parencite{epskamp2018} by estimating group-, subgroup-, and individual-level dynamics in a three-stage process \parencite{Park2024}. First, individual-level results are obtained by estimating a graphical VAR model for each individual. Next, in a procedure similar to that employed by the S-GIMME framework, results from the first stage are used to create an adjacency matrix quantifying structural similarities in estimated dynamics across individuals. However, whereas S-GIMME defines each element of said matrix as the number of shared nonzero paths between each pair of individuals, scGVAR includes shared zero paths in this definition. Following the construction of this similarity matrix, the Walktrap \parencite{Pons2006} algorithm is used to identify subgroups of individuals. These subgroup assignments are then used in the estimation of subgroup-level dynamics, wherein a graphical VAR model is fit to the aggregated time series of individuals in each subgroup. Group-level dynamics are then obtained by fitting a single graphical VAR model to the aggregated time series of all individuals. Similar to S-GIMME, scGVAR is therefore able to accommodate both quantitative and qualitative heterogeneity through the estimation of group-, subgroup-, and individual-level dynamics. However, whereas the S-GIMME procedure allows group- and subgroup-level dynamics to vary in magnitude between individuals, these effects are fixed in the scGVAR approach---that is, all individuals in the same subgroup are characterized by identical parameter estimates. Moreover, the two frameworks differ with respect to the challenges associated with estimation in high-dimensional settings. Indeed, in contrast to the use of fit indices in S-GIMME, scGVAR relies on a penalized estimation procedure to promote sparsity and parsimony.

\section{Multi-VAR}

The motivation for the development of the multi-VAR framework is twofold. First, ILD collected in the social and behavioral sciences are generally composed of data from multiple individuals. From the single-subject VAR perspective, this now involves the estimation of transition matrices for each individual. It is unlikely, however, that the dynamic processes of interest are entirely heterogeneous across all individuals in a given sample. Indeed, a key feature of common ILD modeling approaches, such as multilevel modeling, is that information is shared across individuals in the parameter estimation procedure. A purely idiographic approach, wherein each of the individual transition matrices are estimated independently, disregards this shared information. Thus, the single-subject VAR, though capable of modeling both qualitative and quantitative heterogeneity, is unable to account for shared dynamics and may be suboptimal in many commonly encountered data settings where sharing information across individuals is beneficial (e.g., when time series length is small).

Second, as noted previously, the single-subject VAR is plagued by profligate parameterization \parencite{Sims1980}, such that the number of parameters grows quadratically with each additional component series. Given typical sample sizes (i.e., number of time points) in the social, behavioral, and health sciences \parencite[e.g.,][]{Rot2012}, this could result in the estimation of a large number of parameters given the available data, with such estimates lacking precision \parencite{Bruggemann2012}. Solutions to this dimensionality issue in the single-subject VAR have focused on reducing the parameter space using methods ranging from sequential search procedures to regularization via the least absolute shrinkage and selection operator (Lasso; \cite{Tibshirani1996}). Central to these remedies is the assumption that the true model is sparse in nature. The feasibility of this assumption is open to interpretation; however, the bet on sparsity principle \parencite{Hastie2001} argues that methods assuming sparsity are preferable because if this assumption is false---that is, the true model is dense---then existing approaches are unable to recover the true model without a large amount of data. The multi-VAR framework aims to address the challenges associated with common modeling approaches, such as the single-subject VAR and multilevel modeling, while retaining their desirable features.

To motivate the construction of the multi-VAR framework, we first consider a multivariate ($d$-variate) time series for a single individual, $\{\mathbf{X}_{t}\}_{t \in \mathbb{Z}} = \{(X_{j,t})_{j=1,\dots,d}\}_{t \in \mathbb{Z}}$, where $\mathbf{X}_{t}$ follows the canonical VAR model of order $p$, $\mathrm{VAR}(p)$, if
\begin{equation}
\label{var}
\mathbf{X}_{t} = 
  \boldsymbol{\Phi}_{1} \mathbf{X}_{t-1} + 
  \ldots + 
  \boldsymbol{\Phi}_{p} \mathbf{X}_{t-p} + 
  \mathbf{E}_{t}, \quad t \in \mathbb{Z},
\end{equation}
\noindent for $d \times d$ transition matrices $\boldsymbol{\Phi}_{1}, \dots,\boldsymbol{\Phi}_{p}$ containing autoregressive and cross-regressive parameters and a white noise series $\{\mathbf{E}_{t}\}_{t \in \mathbb{Z}} \sim \text{WN}(\mathbf{0}, \boldsymbol{\Sigma}_{\mathbf{E}})$ characterized by $\mathbb{E}(\mathbf{E}_{t})=0$ and $\mathbb{E}(\mathbf{E}_{t}\mathbf{E}^{'}_{s})=0$ for $s \neq t$. We assume here that $\mathbf{X}_{t}$ is of zero mean for simplicity, though developments that follow can easily accommodate time series with nonzero means. Generally, a unique causal stationary solution to \eqref{var} can be ensured when the roots of $\text{det}( \boldsymbol{\Phi}(z))$, where $\boldsymbol{\Phi}(z)=\mathbf{I}_{d} -\boldsymbol{\Phi}_{1}z - \dots - \boldsymbol{\Phi}_{p} z^{p}$, all have moduli greater than unity. With observations $\mathbf{X}_1, \dots, \mathbf{X}_T$, we can more concisely express \eqref{var} in the familiar regression format:  
\begin{equation}
    \label{bigvar}
    \mathbf{Y}=\boldsymbol{\Phi}\mathbf{Z} + \mathbf{U},
\end{equation}
\noindent where $\mathbf{Y} = (\mathbf{X}_{p+1}, \dots, \mathbf{X}_T)$ is a $d \times (T-p)$ outcome matrix, $\boldsymbol{\Phi} = (\boldsymbol{\Phi}_1, \dots, \boldsymbol{\Phi}_p)$ is a $d \times (dp)$ transition matrix, $\mathbf{Z}$ is a $(dp) \times (T-p)$ design matrix, and $\mathbf{U}$ is a $d \times (T-p)$ matrix of process noise. In the remainder of this work we only consider first-order VAR models---that is, $\mathrm{VAR}(1)$. However, all arguments can be extended to accommodate models with arbitrary lag orders without any loss of generality.

The multi-VAR framework extends the above single-subject VAR representation to readily accommodate multiple-subject, multivariate time series through the estimation of $\boldsymbol{\Phi}^{1}, \dots,\boldsymbol{\Phi}^{K}$ sparse transition matrices for $K$ individuals, each composed of common and unique effects. To do so, the multi-VAR approach relies on the following decomposition of $\boldsymbol{\Phi}^k$,
\begin{equation}
    \label{decomp}
    \boldsymbol{\Phi}^{k} = \boldsymbol{\Gamma} + \boldsymbol{\Upsilon}^{k}, 
    \ k = 1, \dots , K,
\end{equation}
\noindent where $\boldsymbol{\Gamma} \in \mathbb{R}^{d \times d}$ corresponds to the common effects across $K$ individuals and $\boldsymbol{\Upsilon}^{k} \in \mathbb{R}^{d \times d}$ represents the unique effects for individual $k$. This decomposition allows for heterogeneity in the structure of the dynamics through the inclusion of unique, person-specific effects while allowing for some degree of homogeneity in the common effects. Further, as no distributional assumptions are imposed on \eqref{decomp}, individual transition matrices are free to vary both quantitatively and qualitatively across individuals. Notably, shared paths in the common effects matrix are allowed to vary in magnitude---that is, the effects are freely estimated while preserving the structure of $\boldsymbol{\Gamma}$ for all $K$ individuals.

One approach for sparse estimation of \eqref{decomp} was proposed by Fisher and colleagues (\citeyear{Fisher2022}) using the Lasso penalization paradigm,
\begin{equation}
\begin{gathered}
    \label{multivar_obj}
    \underset{\mathbf{\Gamma}, \mathbf{\Upsilon}^1,...,\mathbf{\Upsilon}^K}{\mathrm{argmin}} \frac{1}{N} \sum_{k=1}^K \| \mathbf{Y}^k - (\mathbf{\Gamma} + \mathbf{\Upsilon}^k) \mathbf{Z}^k \|_{2}^2 + P_{standard} \\
    P_{standard} = \lambda_1 \| \mathbf{\Gamma} \|_1 + \sum_{k = 1}^K \lambda_{2,k} \| \mathbf{\Upsilon}^k \|_1,
\end{gathered}
\end{equation}
\noindent where $P_{standard}$ is the standard Lasso penalty, $\| \mathbf{A} \|_1$ denotes the $\ell_1$ norm of $\mathrm{vec(\mathbf{A})}$, and $N = (T - p)$. Sparsity and heterogeneity in the multi-VAR solution is therefore determined and governed by the two penalty parameters, $\lambda_1$ and $\lambda_{2,k}$, which are chosen using cross-validation. The inclusion of penalty parameters on both the common and unique effects allows for flexibility in the approximation of potentially heterogeneous data generating processes. If, for example, individuals share little in common, it would be expected that the solution would be equivalent to estimating independent, single-subject VAR models. This would correspond to large values of $\lambda_1$, such that $\hat{\mathbf{\Gamma}} = \mathbf{0}$ and the solution returns $\hat{\mathbf{\Phi}}^{k} = \hat{\mathbf{\Upsilon}}^{k}$. Conversely, if individuals are highly homogeneous, large values of $\lambda_{2,k}$ would result in $\hat{\mathbf{\Upsilon}}^{k} = \mathbf{0}$ and $\hat{\mathbf{\Phi}}^{k} = \hat{\mathbf{\Gamma}}$, paralleling approaches that pool the time series and estimate a single transition matrix. It is likely that the reality falls between these two extremes, thereby resulting in a solution in which the common and unique effects reflect the degree of heterogeneity observed. 

Despite widespread use of the standard Lasso penalization framework, it is characterized by several important limitations \parencite[for a review, see][]{Freijeiro‐González2022}. Indeed, it is well known that the Lasso exhibits drawbacks with respect to consistent path selection, excessive false positives, and bias in many situations, such as when the variables are strongly dependent. The adaptive Lasso---developed, in part, to address these limitations---replaces the standard $\ell_1$ penalty term with a re-weighted version, where the weights are determined by consistent initial estimates of the model parameters  \parencite{Zou2006}. This weighting enables differential penalization across parameters of interest, such that larger initial estimates correspond to smaller weights (and vice versa for small initial estimates). Using this idea, the adaptive multi-VAR was also proposed by Fisher and colleagues (\citeyear{Fisher2022}), wherein the weights are constructed using initial estimates $\Tilde{\mathbf{\Phi}}^k$ as in
\begin{equation}
    \label{adaptive}
    P_{adaptive} = \lambda_1 \sum_{i,j=1}^d \frac{1}{|\Tilde{\phi}_{i,j,median}|^\alpha}
    |\Gamma_{i,j}| + \sum_{k=1}^K \lambda_{2,k} \sum_{i,j=1}^d
    \frac{1}{|\Tilde{\phi}_{i,j}^k - \Tilde{\phi}_{i,j,median}|^\alpha}|\Upsilon_{i,j}^k|,
\end{equation}
\noindent where $\Gamma_{i,j}$ corresponds to the $\{i,j\}^{th}$ element of $\mathbf{\Gamma}$, $\Upsilon_{i,j}^k$ corresponds to the $\{i,j\}^{th}$ element of $\mathbf{\Upsilon}^k$, and similarly with $\Tilde{\phi}_{i,j}^k$ for $\Tilde{\mathbf{\Phi}}^k$ and $\Tilde{\phi}_{i,j,median}$ for $\Tilde{\mathbf{\Phi}}_{median}$, with $\Tilde{\mathbf{\Phi}}_{median}$ representing the matrix of median coefficient estimates for all $K$ individuals, and $\alpha \geq 1$. Substituting the standard Lasso penalty, $P_{standard}$, in \eqref{multivar_obj} for the adaptive Lasso, $P_{adaptive}$, in \eqref{adaptive} results in the adaptive multi-VAR objective function. Prior work has shown that multi-VAR with adaptive Lasso performs well across a range of factors \parencite{Fisher2024}. 

\section{Subgrouping Multi-VAR}

The subgrouping multi-VAR procedure consists of two steps. First, subgroup enumeration and classification occur. That is, prior to estimation of subgroup-specific effects, the number of relevant subgroups is determined, and each individual is assigned to a subgroup. Next, this information is used in both the construction of the design matrix, $\textbf{Z}$, and the decomposition and penalized estimation of the individual transition matrices, $\mathbf{\Phi}^k$. These two steps are described in detail in the following subsections.

\subsection{Identifying Subgroups}

To identify the number and membership of clusters, subgrouping multi-VAR first estimates $\mathbf{\Phi}^k$ for $K$ individuals using the standard multi-VAR framework. The individual-level effects from these matrices, $\boldsymbol{\Upsilon}^{k}$, are then used to construct a $K \times K$ similarity matrix, where the off-diagonal elements represent the number of shared dynamics between each pair of individuals in terms of presence and sign. If, for example, a specific path is both nonzero and of the same sign (i.e., positive or negative) for individuals $i$ and $j$, then the $\{i,j\}^{th}$ element of the similarity matrix increments by one.  The use of individual-level effects in the construction of the similarity matrix ensures that deviations from common effects contribute to subgroup identification. The construction of the similarity matrix in subgrouping multi-VAR is similar to the procedures implemented in S-GIMME and scGVAR (\cite{Gates2017}; \cite{Park2024}). 

The community detection algorithm Walktrap \parencite{Pons2006} is then applied to this similarity matrix. The Walktrap algorithm uses the information in the similarity matrix to compute a transition matrix, such that each element corresponds to the probability of transitioning from one individual to another for a random walk of a given length. Intuitively, this allows the Walktrap algorithm to identify densely connected areas (i.e., communities). Ward's (\citeyear{Ward1963}) hierarchical clustering procedure is then used to determine the optimal number of clusters by iteratively merging communities until all individuals are in a single cluster. Subgroup enumeration then proceeds by identifying the configuration with the maximum modularity \parencite{Newman2004}, which indicates the degree to which individuals within a cluster are similar relative to those from different clusters. The Walktrap algorithm has been found to perform well when applied to count matrices \parencite{Gates2016}, and has been successfully implemented in similar VAR-based methodological frameworks (e.g., \cite{Gates2017}; \cite{Park2024}).

\subsection{Estimating Subgrouping Effects}

To extend the standard multi-VAR framework to allow for estimation of subgroup-specific effects, we consider the further decomposition of $\mathbf{\Phi}^k$,
\begin{equation}
    \label{subdecomp}
    \boldsymbol{\Phi}^{k} = \boldsymbol{\Gamma} + \boldsymbol{\Pi}^{s} +
    \boldsymbol{\Upsilon}^{k},
    \ s = 1, \dots , S,
    \ k = 1, \dots , K,
\end{equation}
\noindent where $\mathbf{\Gamma}$ and $\mathbf{\Upsilon}^k$ continue to correspond to the common and unique effects, respectively, and $\mathbf{\Pi}^{s} \in \mathbb{R}^{d \times d}$ represents the subgroup effects for subgroup $s$. Paralleling the initial decomposition in \eqref{decomp}, individual transition matrices in \eqref{subdecomp} are free to vary both in structure and magnitude across individuals. This decomposition can be further incorporated into the objective function detailed in \eqref{multivar_obj} within the standard Lasso penalization framework
\begin{equation}
\begin{gathered}
    \label{sub_obj}
    \underset{\mathbf{\Gamma}, \mathbf{\Pi}^1, \dots,\mathbf{\Pi}^S, \dots, \mathbf{\Upsilon}^K}{\mathrm{argmin}} \frac{1}{N} \sum_{k=1}^K \| \mathbf{Y}^k - (\mathbf{\Gamma} + \mathbf{\Pi}^s + \mathbf{\Upsilon}^k) 
    \mathbf{Z}^k \|_{2}^2 + P_{standard} \\
    P_{standard} = \lambda_1 \| \mathbf{\Gamma} \|_1 + 
    \sum_{s = 1}^S \alpha_{s} \| \mathbf{\Pi}^s \|_1 +
    \sum_{k = 1}^K \lambda_{2,k} \| \mathbf{\Upsilon}^k \|_1,
\end{gathered}
\end{equation}
\noindent where the addition of $\alpha_{s}$ in $P_{standard}$ indicates that the sparsity and heterogeneity of the solution is now determined by three penalty parameters, $\lambda_{1}$, $\alpha_{s}$, and $\lambda_{2,k}$. Importantly, this suggests that the competition of the three penalty terms affords even greater flexibility in approximating the underlying data generating processes. In addition to the scenarios noted previously, for example, the subgrouping multi-VAR can accommodate data characterized by within-subgroup homogeneity and a high degree of between-subgroup heterogeneity, resulting in large values of $\lambda_{1}$ and $\lambda_{2,k}$, such that $\hat{\mathbf{\Phi}}^{k} = \hat{\mathbf{\Pi}}^{s}$.

This decomposition can also be incorporated into the penalty function in \eqref{adaptive} for the adaptive multi-VAR
\begin{equation}
\begin{aligned}
    \label{subadaptive}
    P_{adaptive} = & \lambda_1  \sum_{i,j=1}^d \frac{1}{|\Tilde{\phi}_{i,j,median}|^\alpha}
    |\Gamma_{i,j}| \ + \\
    & \sum_{s=1}^S \alpha_{s} \sum_{i,j=1}^d
    \frac{1}{|\Tilde{\phi}_{i,j,median}^s|^\alpha}|\Pi_{i,j}^s| \ + \\
    &
    \sum_{k=1}^K \lambda_{2,k} \sum_{i,j=1}^d
    \frac{1}{|\Tilde{\phi}_{i,j}^k - \Tilde{\phi}_{i,j,median}|^\alpha}|\Upsilon_{i,j}^k|,
\end{aligned}
\end{equation}
\noindent where $\Pi_{i,j}^s$ corresponds to the $\{i,j\}^{th}$ element of $\mathbf{\Pi}^s$, and similarly for $\Tilde{\mathbf{\phi}}^s_{i,j,median}$ and $\Tilde{\mathbf{\Phi}}^s_{median}$. In the current project---as with the standard multi-VAR---the weights in the adaptive Lasso are constructed using initial estimates $\Tilde{\mathbf{\Phi}}^k$, $\Tilde{\mathbf{\Phi}}_{median}$, and $\Tilde{\mathbf{\Phi}}^s_{median}$, with $\Tilde{\mathbf{\Phi}}^s_{median}$ representing the matrix of median coefficient estimates across all individuals in a given subgroup. Note that for the oracle properties of the adaptive Lasso to hold---that is, identification of the correct subset model and optimal estimation rate---consistent estimates for the weights must be chosen \parencite{Zou2006}. In practice, however, selecting an appropriate estimator can be nontrivial in many situations, such as high dimensional settings or when covariates are strongly dependent. In such cases, estimation of adaptive weights via $\ell_1$ or $\ell_2$ penalization, respectively, may convey benefits over standard procedures, such as ordinary least squares or maximum likelihood. For more information regarding the performance and availability of various adaptive weights estimators in the multi-VAR framework, see Fisher et al. (\citeyear{Fisher2024}).

Selection of the unknown penalty parameters---$\lambda_1$, $\alpha_{s}$, and $\lambda_{2,k}$---in the subgrouping multi-VAR framework is done using a blocked cross-validation approach \parencite[BCV;][]{Bulteel2018}, which proceeds as follows. First, each of the $K$ multivariate time series is divided into $F$ equally sized folds. Next, one of the $F$ folds is removed from each time series to serve as the testing block. Using the remaining folds comprising the training block, estimates of $\mathbf{\Gamma}$, $\mathbf{\Pi}^s$, and $\mathbf{\Upsilon}^k$ are obtained, which are subsequently used to predict the testing block and obtain the mean squared error \parencite[MSE; though see][for an alternative to the MSE]{Revol2024}. This procedure continues such that each of the $F$ folds serves as a testing block on which the MSE is computed. The performance is then aggregated across the $K$ individuals and $F$ folds for each combination of $\lambda_1$, $\alpha_{s}$, and $\lambda_{2,k}$, wherein total error is calculated as 
\begin{equation}
    \label{MSE}
    \text{MSE}_{\lambda_1,\alpha_{s},\lambda_{2,k}} =
    \frac{1}{K} \sum_{k=1}^K \frac{1}{F} \sum_{f=1}^F
    \| \hat{\textbf{Y}}_f^k - \textbf{Y}_f^k \|_2^2,
\end{equation}
and the values of the penalty parameters corresponding to the smallest total MSE are selected for the final model. Note that though the subscripts associated with penalty parameters $\alpha_{s}$ and $\lambda_{2,k}$ suggest the possibility of subgroup- and person-specific penalization, the current project only examines a single parameter for each---that is, $\alpha_{s} = \alpha$ and $\lambda_{2,k} = \lambda_{2}$. This more general notation is nonetheless included to emphasize the flexibility of the subgrouping multi-VAR framework.

\section{Simulation Study}

To evaluate the performance of the subgrouping multi-VAR framework, we conducted a Monte Carlo simulation study, wherein factors of interest were selected to represent scenarios typically encountered in ILD applications. Factors examined in the current study included number of individuals, $K = (50, 100)$; length of individual time series, $T = (50, 100)$; number of subgroups, $S = (2, 3)$; and composition of the subgroups, balanced or unbalanced. For each condition, $50$ replications were generated, yielding a fully factorial design with $16$ conditions $(2 \times 2 \times 2 \times 2)$ and $800$ $(16 \times 50)$ unique data sets. In the $K = 50$ conditions, for example, Monte Carlo estimates were computed using 2500 $(50 \times 50)$ multivariate time series. Number of variables was held constant at $d = 10$ for each condition. The performance of the subgrouping extension was assessed in comparison to alternative methods for modeling multi-subject, multivariate time series, namely, S-GIMME and scGVAR (\cite{Gates2017}; \cite{Park2024}). As a benchmark, performance was also compared to the standard multi-VAR framework and multi-VAR with confirmatory subgrouping---that is, subgrouping multi-VAR when subgroup membership is known \textit{a priori}.

The motivation for the selection of the comparison methods, S-GIMME and scGVAR, assessed in the current simulation study was twofold. First, like subgrouping multi-VAR, both frameworks seek to accommodate quantitative and qualitative heterogeneity through the estimation of group-, subgroup-, and individual-level dynamics. Second, both methods address the challenges associated with modeling VAR processes comprised of many variables by inducing sparsity and parsimony in the estimation procedure. As noted previously, however, the manner in which these features are incorporated varies considerably between approaches. For example, whereas subgrouping multi-VAR estimates group-, subgroup-, and individual-level dynamics simultaneously, S-GIMME and scGVAR are characterized by iterative and stage-based estimation procedures. In addition to assessing the performance of subgrouping multi-VAR across a range of factors, the current simulation study therefore represents a rigorous evaluation of competing approaches for the analysis of multiple-subject, multivariate time series. 

\subsection{Data Generation and Model Estimation}

Across all conditions, $10 \times 10$ transition matrices were generated for each individual in the following manner. First, subgroup membership was specified. For the balanced conditions, subgroups were constructed such that each contained the same number of individuals. For the unbalanced conditions with two subgroups, 30\% of individuals were placed into the first subgroup, and the remaining 70\% were specified as members of the second subgroup. For the unbalanced conditions with three subgroups, the first and second subgroups consisted of 20\% of individuals, and the third subgroup contained the remaining 60\%. Next, common effects, $\mathbf{\Gamma}$, were specified as the 10 diagonal elements of each transition matrix---that is, the autoregressive effects. The location of subgroup, $\mathbf{\Pi}^{s}$, and unique, $\mathbf{\Upsilon}^{k}$, effects were then chosen at random, wherein the number of effects for each represented 5\% of possible paths. Thus, each transition matrix consisted of 10 common effects, five subgroup effects, and five unique effects, yielding a sparse matrix with 20\% nonzero entries. All effects were drawn from a $\mathcal{U}(0,1)$ distribution until a stationary solution was obtained, as determined by the stationarity condition noted previously. The data generation procedure therefore incorporated both qualitative and quantitative heterogeneity through variation in the location of the subgroup and unique effects and variation in the magnitude of group, subgroup, and unique effects, respectively. 

All models analyzed within the multi-VAR framework---that is, subgrouping multi-VAR, multi-VAR with confirmatory subgrouping, and standard multi-VAR---were estimated using the adaptive Lasso procedure detailed above, with adaptive weights obtained via $\ell_1$ penalization. Paralleling the group-level stage in the S-GIMME framework, however, all autoregressive paths were assumed to be nonzero, and were therefore not subjected to penalization in either the subgroup enumeration or estimation stages. All multi-VAR and S-GIMME analyses were conducted using the \texttt{multivar} \parencite{multivar} and \texttt{gimme} \parencite{gimme} R packages, respectively.

\subsection{Outcome Measures}

The performance of the subgrouping multi-VAR framework and comparison methods was evaluated in three ways: model recovery, quality of estimated effects, and accuracy of subgroup identification. A number of metrics were used to assess model recovery, including sensitivity, specificity, and Matthews correlation coefficient \parencite[MCC;][]{Matthews1975}. Sensitivity and specificity can be interpreted as the true positive and true negative rates, respectively, and were computed as follows: 
\begin{equation}
    \label{sens}
    \text{Mean sensitivity} = \frac{1}{K} \sum_{k = 1}^{K} 
    \left( \frac{ \sum_{i,j} (\hat{\phi}_{i,j}^k \neq 0 \text{ and } \phi_{i,j}^k \neq 0 ) }
    { \sum_{i,j} (\phi_{i,j}^k \neq 0) } \right),
\end{equation}
\begin{equation}
    \label{spec}
    \text{Mean specificity} = \frac{1}{K} \sum_{k = 1}^{K} 
    \left( \frac{ \sum_{i,j} (\hat{\phi}_{i,j}^k = 0 \text{ and } \phi_{i,j}^k = 0 ) }
    { \sum_{i,j} (\phi_{i,j}^k = 0) } \right),
\end{equation}
where $\hat{\phi}_{i,j}^k$ and $\phi_{i,j}^k$ correspond to the $\{i,j\}^{th}$ element of the estimated and true transition matrices for the $k^{th}$ individual, respectively. Mean sensitivity and specificity were then averaged across all replications, resulting in sensitivity and specificity values for each condition. MCC similarly provides a metric for assessing the recovery of the data generating model by incorporating both sensitivity and specificity in its calculation:
\begin{equation}
    \label{MCC}
    \text{Mean MCC} = \frac{1}{K} \sum_{k = 1}^K
    \left( \frac{TP_k \times TN_k - FP_k \times FN_k}
    {\sqrt{(TP_k + FP_k)(TP_k + FN_k)(TN_k + FP_k)(TN_k + FN_k)}} \right),
\end{equation}
where $TP$ is the number of parameters correctly estimated as nonzero, $TN$ is the number of parameters correctly estimated as zero, $FP$ is the number of parameters incorrectly estimated as nonzero, and $FN$ is the number of parameters incorrectly estimated as zero. Notably, MCC ranges from perfect disagreement between estimated and true models ($MCC = -1$) to perfect agreement between the two ($MCC = 1$), and can therefore be interpreted as a discretization of the standard correlation coefficient in the context of binary classification \parencite{Boughorbel2017}. As such, the MCC can be interpreted with respect to commonly employed benchmarks \parencite[e.g.,][]{Cohen1988}.

To evaluate the quality and variability of the estimated parameters, we computed the mean absolute bias and root mean squared error (RMSE):
\begin{equation}
    \label{MAB}
    \text{Mean Absolute Bias} = \frac{1}{K} \sum_{k=1}^K \frac{1}{d^2} \sum_{i,j=1}^d
    |\hat{\phi}_{i,j}^k - \phi_{i,j}^k|,
\end{equation}
\begin{equation}
    \label{RMSE}
    \text{RMSE} = \frac{1}{K} \sum_{k=1}^K \sqrt{\frac{1}{d^2} \sum_{i,j=1}^d
    (\hat{\phi}_{i,j}^k - \phi_{i,j}^k)^2},
\end{equation}
where $\hat{\phi}_{i,j}^k$ and $\phi_{i,j}^k$ again correspond to the $\{i,j\}^{th}$ element of the estimated and true transition matrices for the $k^{th}$ individual, respectively, and $d$ represents the number of variables (i.e., 10). These values were then averaged over all replications to provide mean absolute bias and RMSE metrics for each condition.

To assess the accuracy of subgroup identification, we computed the Hubert-Arabie adjusted Rand index \parencite[ARI;][]{Hubert1985}:
\begin{equation}
    \label{ARI}
    \text{ARI} = \frac{{K \choose 2} (a+d) - [(a+b)(a+c) + (c+d)(b+d)]}
    {{K \choose 2}^2 - [(a+b)(a+c) + (c+d)(b+d)]},
\end{equation}
where $a$ represents the number of pairs of individuals correctly placed in the same cluster, $b$ is the number of pairs incorrectly placed in different in different cluster, $c$ indicates the number of pairs incorrectly placed in the same cluster, and $d$ is the number of pairs correctly placed in different clusters. The ARI therefore incorporates information about the number of true positive, false negative, false positive, and true negative classifications. The ARI has an upper bound of 1, which corresponds to perfect subgroup identification. ARI values of 0, conversely, indicate that subgroup assignments were equal to chance. ARI values greater than or equal to 0.90 were considered excellent, values between 0.80 and 0.89 were considered good, values between 0.65 and 0.79 indicated moderate recovery, and values below 0.64 suggested poor subgroup identification \parencite{Steinley2004}. Monte Carlo errors (MCE), defined as the standard deviation of the Monte Carlo estimates across all replications, were computed to quantify the uncertainty in all outcome measure estimates \parencite{Koehler2009}.

\subsection{Simulation Results}

\paragraph{Model Recovery}

\begin{figure}
    \centering
    \includegraphics{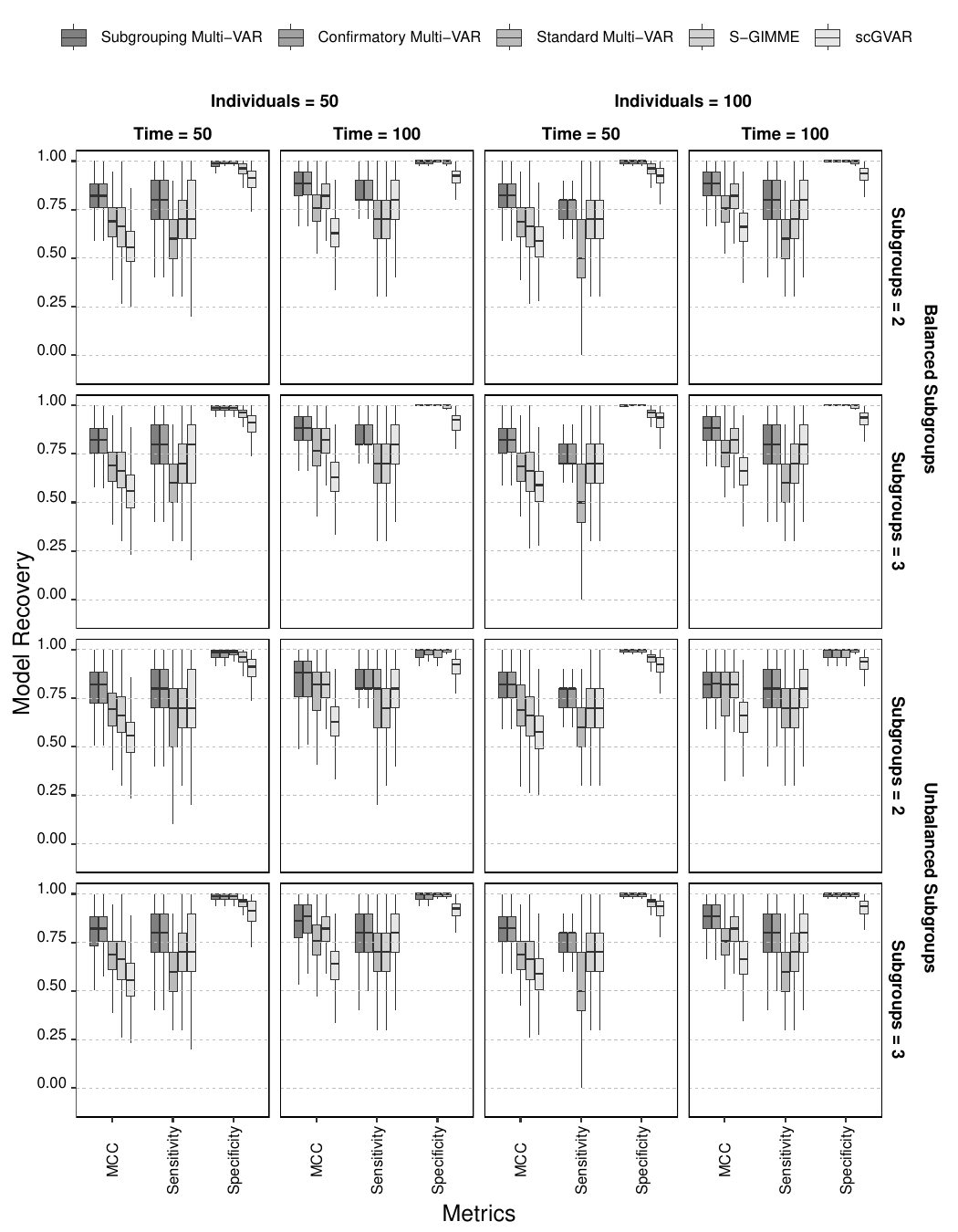}
    \caption{Model recovery metrics for each condition and model.}
    \label{fig:recovery}
\end{figure}

Sensitivity, specificity, and MCC values across all conditions for each estimator can be seen in Tables \ref{tab:sens_spec} and \ref{tab:mcc_ari}, and are visualized in Figure \ref{fig:recovery}. Given that MCC incorporates both sensitivity and specificity, assessment of model recovery focused on this metric. Across all conditions examined in the current study, subgrouping multi-VAR demonstrated more accurate model recovery than S-GIMME, scGVAR, or standard multi-VAR, and displayed only a slight decrement in performance when compared to the confirmatory subgrouping framework, where the true subgroup membership is assumed to be known. When $K = 50$ and $S = 2$, mean MCC for subgrouping multi-VAR increased as a function of time series length, with values of 0.81 and 0.88 when $T = 50$ and $T = 100$, respectively, for balanced subgroups, and values of 0.79 and 0.84 when subgroup membership was unbalanced. Subgrouping multi-VAR therefore demonstrated good model recovery for these conditions, with some MCC values reflecting near perfect agreement between true and estimated models. Differences in model recovery between multi-VAR with data-driven subgrouping and confirmatory subgrouping were minimal across these conditions (less than 0.02). Model recovery for S-GIMME, scGVAR, and standard multi-VAR similarly increased as a function of time series length. Mean MCC values for standard multi-VAR were 0.68 and 0.77 when $T = 50$ and $T = 100$, respectively, when subgroups were balanced, and 0.69 and 0.78 when subgroups were unbalanced. Mean MCC values for S-GIMME were 0.66 and 0.80 for both balanced and unbalanced subgroups, whereas values for scGVAR were 0.56 and 0.64 for balanced subgroups and 0.55 and 0.63 for unbalanced subgroups. 

Similar results were observed when $K = 50$ and $S = 3$. Indeed, mean MCC values for subgrouping multi-VAR increased from 0.80 to 0.88 as $T$ increased from 50 to 100 for subgroups with balanced membership, and from 0.78 to 0.84 for unbalanced subgroups. Observed results therefore suggest that model recovery for subgrouping multi-VAR was largely unaffected by the addition of a third subgroup when $K = 50$. Discrepancies between data-driven and confirmatory subgrouping approaches were again minimal across these conditions. Mean MCC values for standard multi-VAR were 0.68 and 0.78 when subgroups were balanced, and 0.68 and 0.76 when membership was unbalanced. Comparable metrics were observed for both S-GIMME and scGVAR. Thus, model recovery for subgrouping multi-VAR and comparison approaches was largely unaffected by the inclusion of an additional subgroup when $K = 50$.

When $K = 100$ and $S = 2$, mean MCC values for subgrouping multi-VAR similarly increased as a function of time series length, with values of 0.82 when $T = 50$ and 0.88 for $T = 100$ when subgroups were balanced, and values of 0.79 and 0.82 when unbalanced. The addition of a third balanced subgroup resulted in similar model recovery metrics for subgrouping multi-VAR, with mean MCC values 0.81 and 0.87. Notably, model recovery was better for three unbalanced subgroups than two unbalanced subgroups, with mean MCC values of 0.81 when $T = 50$ and 0.86 when $T = 100$. In general, observed results suggest that increasing $K$ from 50 to 100 resulted in a slight improvement in model recovery. Discrepancies between data-driven and confirmatory approaches remained minimal. Model recovery for S-GIMME mirrored results observed in previously reported conditions, with mean MCC values of 0.66 and 0.80 when $T = 50$ and $T = 100$, respectively, for each combination of number of subgroups and membership composition. Conversely, model recovery for scGVAR improved when $K = 100$, with mean MCC values ranging from 0.58 to 0.59 when $T = 50$ and from 0.66 to 0.67 when $T = 100$. Standard multi-VAR displayed a slight decrease in model recovery when $K = 100$, with mean MCC values of approximately 0.67 for $T = 50$ and 0. 75 for $T = 100$ across all subgroup number and membership composition conditions.

\paragraph{Parameter Estimates}

\begin{figure}
    \centering
    \includegraphics{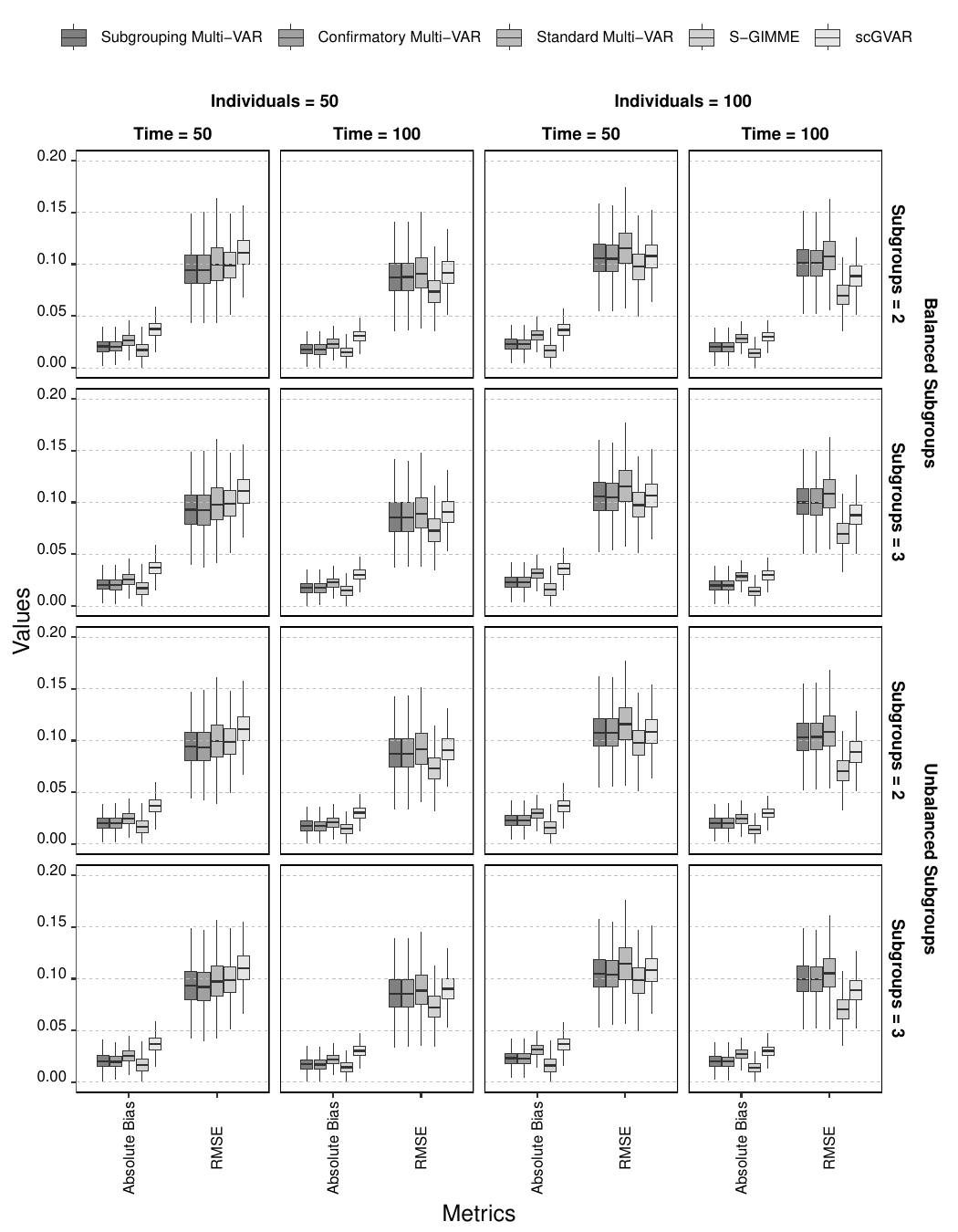}
    \caption{Quality and variability of parameter estimates for each condition and model.}
    \label{fig:quality}
\end{figure}

Mean absolute bias and RMSE values can be found in Table \ref{tab:bias_rmse} and are visualized in Figure \ref{fig:quality}. Mean absolute bias for subgrouping multi-VAR was largely consistent across all conditions examined in the current study, with values ranging from 0.018 to 0.023. In general, bias was at a minimum when $K = 50$ and $T = 100$, mirroring results observed for model recovery. Conversely, bias was largest when $K = 100$ and $T = 50$, suggesting that an increase in sample size without a corresponding increase in time series length results in reduced performance. Thus, large sample sizes may result in a high degree of heterogeneity, thereby diminishing model performance when the number of time points is small \parencite[e.g.,][]{Fisher2024}. It is worth noting, however, that these differences were small in magnitude, and may be a function of sampling variability, consistent with both observed MCE estimates and visualization of estimates in Figure \ref{fig:quality}. There were no meaningful differences observed between data-driven and confirmatory approaches with respect to mean absolute bias. Differences were observed, however, between subgrouping and standard multi-VAR, such that failing to account for the presence of subgroups increased mean absolute bias across all conditions, with values ranging from 0.021 to 0.032. Notably, compared to subgrouping multi-VAR, mean absolute bias for S-GIMME estimates was both smaller and less variable across all conditions, with values ranging from 0.014 to 0.018. Mean absolute bias for S-GIMME was smallest when $K = 100$ and $T = 100$, and largest when $K = 50$ and $T = 100$. Conversely, mean absolute bias for scGVAR estimates was larger than that observed for subgrouping multi-VAR across all conditions, with values ranging from 0.030 to 0.037.

A similar pattern of results was observed for the RMSE of subgrouping multi-VAR estimates. Indeed, RMSE was at a minimum when $K = 50$ and $T = 100$, and at a maximum when $K = 100$ and $T = 50$, with values ranging from 0.087 to 0.109 across all conditions. Differences in RMSE values between data-driven and confirmatory subgrouping approaches were less than 0.01 across all conditions. RMSE values for standard multi-VAR were larger across all conditions, with values ranging from 0.091 to 0.117. RMSE values for S-GIMME estimates were larger than those observed for subgrouping multi-VAR when both $K$ and $T$ were at a minimum, and smaller for all other conditions. Thus, the subgrouping multi-VAR estimates were less variable than S-GIMME estimates when both sample size and time series length were small. In general, subgrouping multi-VAR estimates were less variable than scGVAR estimates. However, RMSE values for scGVAR estimates were smaller than values for subgrouping multi-VAR estimates when both $K$ and $T$ were at a maximum.

\paragraph{Subgroup Recovery}

ARI values for subgrouping multi-VAR, S-GIMME, and scGVAR are shown in Table \ref{tab:mcc_ari}. In general, subgroup recovery was considered good or excellent for subgrouping multi-VAR across all conditions \parencite{Steinley2004}. Subgroup recovery was at a maximum (ARI = 0.98) when $K = 50$, $T = 100$, and subgroup composition was balanced, indicating near perfect subgroup recovery. Subgroup recovery was at a minimum when $K = 100$, $T = 50$, with three unbalanced subgroups (ARI = 0.88). Conversely, subgroup recovery for S-GIMME was considered poor across all conditions, with ARI values ranging from 0.01 to 0.64. The minimum ARI value for S-GIMME was observed when $K = 50$, $T = 50$, with two unbalanced subgroups, and the maximum value occurred when $K = 100$, $T = 100$, with two balanced subgroups. Subgroup recovery for scGVAR ranged from moderate to excellent across all conditions, such that ARI values were at a minimum of 0.73 when $K = 50$, $T = 50$, with three unbalanced subgroups, and a maximum of 0.98 when $K = 100$, $T = 100$, with two subgroups. 

Mirroring observed ARI values, subgrouping multi-VAR generally recovered the correct number of subgroups. Observed differences in ARI values between subgrouping multi-VAR and S-GIMME were similarly consistent with discrepancies in the number of subgroups identified by both frameworks. When $K = 100$ and $T = 50$, for example, the number of subgroups recovered ranged from 12 to 17 for S-GIMME and three to six for subgrouping multi-VAR. Thus, though both approaches overestimated the number of subgroups in these conditions, S-GIMME was adversely impacted to a greater degree. This is consistent with prior work showing that recovery of nonzero dynamics in GIMME is reduced when the number of time points is small \parencite{Nestler2021}. Notably, scGVAR subgroup enumeration was not affected by time series length, suggesting that it may be particularly well-suited to settings in which number of time points is small. 

\section{Empirical Example}

To demonstrate the utility of the subgrouping multi-VAR framework, we present an empirical example using data from Fisher et al. (\citeyear{Fisher2017}). Data consisted of 40 individuals with a primary diagnosis of either major depressive disorder (MDD) or generalized anxiety disorder (GAD) who were assessed four times per day for 30 days. For the purposes of the current application, we restricted our analyses to the 10 variables related to MDD (e.g., \textit{down and depressed}) and GAD (e.g., \textit{worried}) symptomatology, thereby ensuring that the number of variables examined mirrored those assessed in the simulation study. All variables were measured using a visual analogue scale ranging from $0-100$, such that 0 indicated that the participant did not experience the symptom at all during the preceding hours and 100 indicated that the symptom was experienced as much as possible. As the multi-VAR framework does not currently accommodate missing data, linear imputation via the \texttt{imputeTS} package \parencite{imputeTS} was employed. For more information regarding data characteristics and study procedures, see Fisher et al. (\citeyear{Fisher2017}).

The motivation for the selection of the current empirical dataset was influenced by several factors. First, both the number of individuals and length of time series were consistent with data analyzed in the simulation study. The quality of the group-, subgroup-, and individual-level estimates were therefore unlikely to be impacted by characteristics of the dataset, such as sample size. Second, the symptomatology of many psychopathological syndromes, including MDD and GAD, is characterized by persistent heterogeneity \parencite[e.g.,][]{Kotov2017}. Indeed, prior work identified 1030 unique symptom profiles in a sample of individuals diagnosed with MDD \parencite{Fried2015}. Thus, the current example represents an evaluation of the degree to which subgrouping multi-VAR accommodates quantitative and qualitative heterogeneity in an empirical setting. Finally, all participants had a primary diagnosis of either MDD or GAD. Comparisons between diagnostic status obtained via structured clinical interview and subgroup membership derived by subgrouping multi-VAR were therefore possible. 

Estimation of group-, subgroup-, and individual-level effects proceeded in the manner detailed above. First, standard multi-VAR without subgrouping was fit to the data. Estimates of the individual-level effects, $\hat{\boldsymbol{\Upsilon}}^k$, were then used to derive subgroup membership. Using these subgroup labels, subgrouping multi-VAR was then fit to the data. Paralleling the modeling procedure for the simulated data, the autoregressive effects were not subjected to penalization via the adaptive Lasso in either the subgroup enumeration or estimation stages. In addition to analysis of the estimated group-, subgroup-, and individual-level effect, results were examined with respect to several between-person variables. That is, associations between subgroup membership and constructs of interest, such as diagnostic status, were assessed. All analyses were conducted using the \texttt{multivar} R package \parencite{multivar}.

\subsection{Empirical Results}

\begin{figure}
    \centering
    \includegraphics{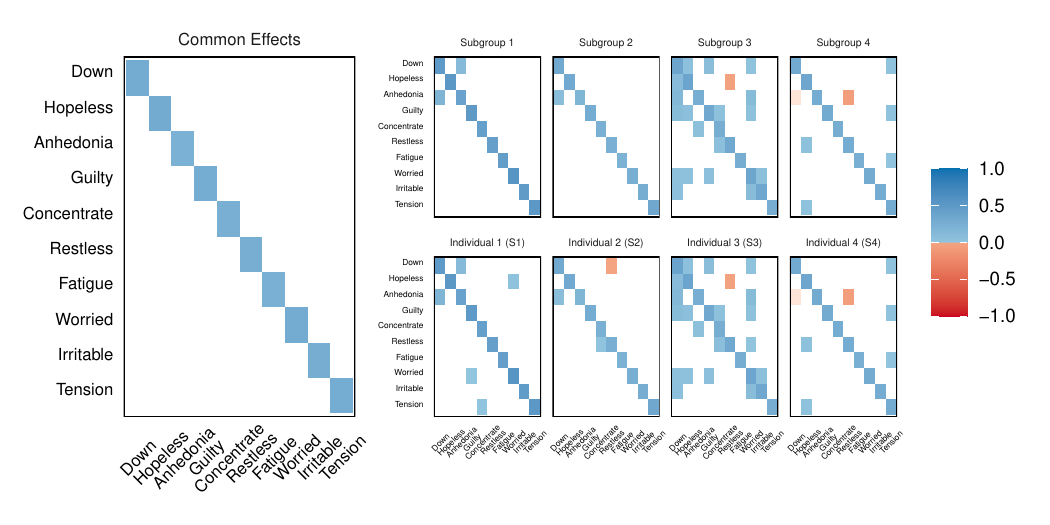}
    \caption{Left: common effects across all individuals. Right: subgroup (e.g., Subgroup 1) and total (e.g., Individual 1 (S1)) effects for four representative individuals.}
    \label{fig:empirical}
\end{figure}

Estimated effects are depicted in Figure \ref{fig:empirical}. Estimates of common effects consisted of all autoregressive paths, with magnitudes ranging from 0.23 to 0.31. This autoregressive behavior is consistent with findings from the original study \parencite{Fisher2017}. With respect to subgroup enumeration, subgrouping multi-VAR identified four subgroups. The first three subgroups consisted of 25, 10, and four individuals, respectively, and the fourth subgroup was defined by a single individual. Given its singleton status, the fourth subgroup was excluded from subsequent analyses. Estimated effects for the first subgroup consisted of all autoregressive paths, with magnitudes ranging from 0.17 to 0.26, and a bidirectional relationship between anhedonia and down and depressed. The second subgroup was also characterized by estimated autoregressive effects, though most of these were negative, indicating that participants in this subgroup had autoregressive dynamics that were smaller in magnitude than those estimated at the group level. A relation between down and depressed and anhedonia was also observed for participants in this subgroup, though it was unidirectional in nature. Thus, the first two subgroups were distinguished by differences in both the sign of the estimated autoregressive effects and the nature of the dynamic relation between down and depressed and anhedonia.

The third subgroup also featured estimates of all autoregressive effects, with magnitudes ranging from $-0.01$ to $0.08$, suggesting that, in general, participants in this subgroup were characterized by slightly stronger autoregressive dynamics than those derived at the group level. Notably, differences between this subgroup and the first two were observed with respect to the number of estimated cross-lagged effects. Indeed, out of 90 possible cross-lagged paths, 19 were estimated as nonzero, suggesting that the network of dynamic relations between MDD and GAD symptoms was much denser for individuals in this subgroup. Despite the observed qualitative and quantitative heterogeneity in estimated subgroup effects, subgroup membership was not associated with diagnostic status or between-person measures of depression and anxiety. However, given the degree of comorbidity in the present sample---75\% of individuals had at least one comorbid diagnosis---identifying clear distinctions between participants with respect to diagnostic status and symptom level is a challenging endeavour.

Estimates of individual-level effects varied both qualitatively and quantitatively between individuals. For some individuals, these effects corresponded to the autoregressive paths, thereby serving to increase or decrease the magnitude of these dynamics, depending on the sign, compared to group- and subgroup-level estimates. Estimated effects for other individuals epitomized the persistent qualitative heterogeneity inherent in many dynamic processes of interest, such that nonzero cross-lagged paths represented an increase in both the complexity and density of symptom relations. Notably, estimates of individual-level effects for one individual were all zero, suggesting that their pattern of dynamics was fully captured by the group- and subgroup-level estimates. 

\section{Discussion}

Many dynamic processes are characterized by persistent heterogeneity in both the magnitude of the effects of interest and the functional form of the process itself. Failing to appropriately account for such heterogeneity when present limits the degree to which observed results can be generalized across various levels of analysis \parencite{Molenaar2004}. The current paper sought to address this challenge through the introduction of a subgrouping extension to the multi-VAR framework. Built on the VAR model, subgrouping multi-VAR is a method for analyzing multiple-subject, multivariate time series that incorporates both data-driven subgroup identification and penalized estimation of group-, subgroup-, and individual-level effects. In contrast to similar approaches, the incorporation of penalty parameters at each level of analysis allows for both sparsity in the estimated transition matrices and flexible approximation of potentially heterogeneous data generating processes. In doing so, it helps to bridge the divide between nomothetic and idiographic approaches to the analysis of ILD. The efficacy and utility of the subgrouping multi-VAR approach was demonstrated in a simulation study and empirical example.

The simulation study assessed the performance of the subgrouping multi-VAR framework across a range of design factors. Moreover, the current approach was compared to two alternative multiple-subject, multivariate time series methods, as well as benchmark comparisons with the standard multi-VAR framework and multi-VAR with confirmatory subgrouping (i.e., when true subgroup membership was known). With respect to model recovery, subgrouping multi-VAR outperformed S-GIMME, scGVAR, and the standard multi-VAR approach across all conditions, and demonstrated only a slight decrease in performance compared to multi-VAR with confirmatory subgrouping. Model recovery was strongest for the condition corresponding to 50 individuals, 100 time points, and three balanced subgroups, and weakest for 50 individuals, 50 time points, and three unbalanced subgroups. Even in this weakest condition, however, subgrouping multi-VAR demonstrated good recovery of the true data generating process. Observed differences in model recovery between subgrouping multi-VAR and comparison methods are likely a function of both the unique manner in which the current approach models heterogeneity in dynamic processes and the quality of the estimates used in the subgroup enumeration stage. Indeed, the structured nature of the penalization procedure and simultaneous estimation of effects of interest ensures that the heterogeneity of the solution is determined by the competition of the three penalty parameters corresponding to the group-, subgroup-, and individual-level effects. Conversely, estimation of heterogeneous dynamics in the S-GIMME framework proceeds iteratively and is a function of user-specified thresholds, modification indices, and fit statistics. The scGVAR approach is similarly iterative in nature, such that effects at different levels of analysis are estimated separately. Though prior work has shown that both S-GIMME and scGVAR demonstrate good model recovery (e.g., \cite{Gates2017}; \cite{Park2024}), the subgrouping multi-VAR estimation procedure described herein may better approximate data generating processes characterized by a high degree of quantitative and qualitative heterogeneity, consistent with observed results. 

Notably, this was also observed in the subgroup enumeration stage, such that subgrouping multi-VAR exhibited excellent identification of subgroup membership across nearly all conditions. Thus, standard multi-VAR estimates used in the construction of the Walktrap adjacency matrix \parencite{Pons2006} can effectively inform the derivation of shared patterns of dynamics. These results contrast those observed for S-GIMME, wherein subgroup recovery was considered poor across all conditions. Indeed, S-GIMME often overestimated the number of subgroups, whereas subgrouping multi-VAR and scGVAR more consistently recovered the correct subgroup structure. These results therefore suggest that subgroup enumeration in the S-GIMME framework, wherein group-stage estimates are used to derive subgroup membership, is less effective than approaches that employ individual-level estimates, such as subgrouping multi-VAR and scGVAR. It is worth noting, however, that observed differences in subgroup identification between subgrouping multi-VAR and comparison methods may not generalize to situations in which subgroup-level effects are driven by contemporaneous dynamics, as the multi-VAR framework does not accommodate such dynamics. Prior S-GIMME work, for example, observed subgroup-level effects that consisted entirely of directed contemporaneous relations \parencite{Lane2019}. Future work should examine the behavior of subgrouping multi-VAR under such conditions.

Mirroring model recovery results, absolute bias and RMSE for subgrouping multi-VAR estimates were lowest for conditions corresponding to 50 individuals, 100 time points, and three subgroups. Largest values of absolute bias and RMSE were observed for conditions with 100 individuals and 50 time points, suggesting that analyses characterized by a large number of individuals may require a correspondingly large number of time points. In general, however, the quality and variability of estimated parameters---as indicated by absolute bias and RMSE---remained relatively stable across conditions. More work is needed to determine the behavior of these metrics as the discrepancy between number of individuals and number of time points increases. In contrast to model recovery results, absolute bias and RMSE values for subgrouping multi-VAR estimates were not consistently preferable when compared those obtained for S-GIMME estimates. Across all conditions, for example, subgrouping multi-VAR parameter estimates exhibited a greater degree of bias than S-GIMME estimates, though these differences were small in magnitude. Similar results were observed for RMSE values when data consisted of either 100 individuals or 100 time points. However, when both number of time points and number of individuals were small, multi-VAR parameter estimates were less variable than those obtained from S-GIMME. Observed differences in the quality and variability of parameter estimates between subgrouping multi-VAR and S-GIMME are generally consistent with prior work showing that GIMME estimates exhibit lower absolute bias and RMSE values than standard multi-VAR when the data are highly heterogeneous \parencite{Fisher2024}. Differences in absolute bias and RMSE values between subgrouping multi-VAR and scGVAR were more consistent with model recovery results. That is, subgrouping multi-VAR estimates were less biased and variable than scGVAR estimates across most conditions. 

Results from an empirical example using data from Fisher et al. (\citeyear{Fisher2017}) demonstrated the utility of the subgrouping multi-VAR framework. At the group level, symptom relations were characterized entirely by autoregressive dynamics, and were consistent with results observed in other clinical samples (e.g., \cite{DeVos2017}; \cite{Park2024}). Symptoms of MDD and GAD therefore exhibited some degree of inertia \parencite[e.g.,][]{Kuppens2010} for all individuals. In addition to group-level effects, three primary subgroups were identified, each distinguished by differences in both the magnitude and structure of symptom dynamics. Though subgroup membership was not associated with between-person variables of interest, such as diagnostic status, identification of shared patterns of symptom relations is not without utility. Indeed, observed findings are consistent with prior work showing that meaningful patterns of subgroup-specific dynamics are not necessarily associated with mean levels of symptoms \parencite{Lane2019}. Such dynamics could, for example, inform future research and treatment efforts, consistent with the recent shift in focus toward the development of personalized models of psychopathology \parencite{Wright2020}. The degree of quantitative and qualitative heterogeneity observed with respect to estimated individual-level effects similarly indicates a need for person-specific approaches to prevention and intervention \parencite[e.g.,][]{Fisher2019}.

Simulation and empirical results observed in the current study should be interpreted in the context of several limiting factors, each of which represents an area for future research. First, as noted above, the multi-VAR framework cannot estimate contemporaneous dynamics. As such, subgrouping multi-VAR may not be an appropriate methodological approach for processes characterized primarily by said dynamics. Next, despite work showing that Walktrap is a reliable community detection algorithm \parencite{Gates2016}, the efficacy of alternative clustering approaches was not assessed. Recent work, for example, suggests that Walktrap could be improved by replacing Ward's hierarchical clustering algorithm \parencite{Ward1963} with methods for \textit{K}-means clustering \parencite{Brusco2024}. Moreover, approaches relying on modularity optimization may fail to identify subgroups consisting of a small number of individuals \parencite{Fortunato2007}. More work is needed to determine if alternative approaches improve the performance of subgrouping multi-VAR. Finally, observed results may not generalize to conditions not assessed in the current simulation and empirical studies. Indeed, prior work has demonstrated that model recovery for standard multi-VAR is impacted by the number of variables comprising the multivariate time series, such that performance improves as number of variables increases \parencite{Fisher2022}. 

The current study introduced a novel methodological framework for analyzing multiple-subject, multivariate time series characterized by persistent quantitative and qualitative heterogeneity. Results from both an extensive simulation study and empirical example suggest that subgrouping multi-VAR is an effective approach for estimating group-, subgroup-, and individual-level dynamics. Notably, the current approach demonstrated good model recovery under commonly encountered conditions, such as when both sample size and time series length are small. Moreover, the advantage of subgrouping multi-VAR over popular alternatives with respect to model recovery was greatest when time series length was small, suggesting that the current approach may be particularly well suited for the types of data frequently collected in the social, behavioral, and health sciences.

\begin{table}[ht]
\centering
\caption{Sensitivity and Specificity}
\label{tab:sens_spec}
\scalebox{0.5}{
\begin{threeparttable}
\begin{tabular}{rrrrcccccccccc}
  \toprule
  &&&& \multicolumn{10}{c}{Measure}\\
 \cmidrule(lr){5-14} &&&& \multicolumn{5}{c}{Sensitivity} &
    \multicolumn{5}{c}{Specificity} \\
 \cmidrule(lr){5-9} \cmidrule(lr){10-14} &&&& \multicolumn{5}{c}{Model} & 
    \multicolumn{5}{c}{Model}\\
 \cmidrule(lr){5-9} \cmidrule(lr){10-14} \multicolumn{1}{c}{Time} &
    \multicolumn{1}{c}{Subjects} &
    \multicolumn{1}{c}{Subgroups} &
    \multicolumn{1}{c}{Balance} &
    \multicolumn{1}{c}{multi-VAR (S)} &  
    \multicolumn{1}{c}{multi-VAR (CS)} &
    \multicolumn{1}{c}{multi-VAR} &
    \multicolumn{1}{c}{S-GIMME} &
    \multicolumn{1}{c}{scGVAR} &
    \multicolumn{1}{c}{multi-VAR (S)} &  
    \multicolumn{1}{c}{multi-VAR (CS)} &
    \multicolumn{1}{c}{multi-VAR} &
    \multicolumn{1}{c}{S-GIMME} &
    \multicolumn{1}{c}{scGVAR} \\
    \midrule
  50 &  50 &   2 & Balanced & 0.77 (0.14) & 0.78 (0.13) & 0.59 (0.18) & 0.71 (0.15) & 0.73 (0.17) & 0.99 (0.02) & 0.99 (0.01) & 0.99 (0.02) & 0.96 (0.03) & 0.90 (0.06) \\ 
  50 &  50 &   2 & Unbalanced & 0.78 (0.14) & 0.78 (0.13) & 0.64 (0.18) & 0.71 (0.15) & 0.73 (0.17) & 0.98 (0.02) & 0.98 (0.02) & 0.98 (0.02) & 0.96 (0.03) & 0.90 (0.07) \\ 
  50 &  50 &   3 & Balanced & 0.77 (0.15) & 0.77 (0.13) & 0.59 (0.18) & 0.71 (0.15) & 0.74 (0.17) & 0.99 (0.02) & 0.99 (0.02) & 0.99 (0.02) & 0.96 (0.03) & 0.90 (0.07) \\ 
  50 &  50 &   3 & Unbalanced & 0.75 (0.16) & 0.78 (0.13) & 0.60 (0.19) & 0.71 (0.15) & 0.73 (0.17) & 0.98 (0.02) & 0.99 (0.02) & 0.99 (0.02) & 0.96 (0.03) & 0.90 (0.07) \\ 
  50 & 100 &   2 & Balanced & 0.75 (0.16) & 0.76 (0.13) & 0.53 (0.20) & 0.71 (0.15) & 0.72 (0.18) & 0.99 (0.01) & 1.00 (0.01) & 1.00 (0.01) & 0.96 (0.03) & 0.92 (0.06) \\ 
  50 & 100 &   2 & Unbalanced & 0.74 (0.15) & 0.75 (0.13) & 0.57 (0.19) & 0.71 (0.15) & 0.72 (0.17) & 0.99 (0.02) & 0.99 (0.02) & 0.99 (0.02) & 0.96 (0.03) & 0.92 (0.06) \\ 
  50 & 100 &   3 & Balanced & 0.73 (0.16) & 0.74 (0.13) & 0.52 (0.19) & 0.71 (0.15) & 0.72 (0.18) & 0.99 (0.01) & 1.00 (0.01) & 1.00 (0.01) & 0.96 (0.03) & 0.92 (0.06) \\ 
  50 & 100 &   3 & Unbalanced & 0.74 (0.17) & 0.76 (0.13) & 0.53 (0.20) & 0.71 (0.15) & 0.71 (0.17) & 0.99 (0.01) & 0.99 (0.01) & 0.99 (0.01) & 0.95 (0.03) & 0.92 (0.06) \\ 
  100 &  50 &   2 & Balanced & 0.82 (0.12) & 0.83 (0.11) & 0.66 (0.16) & 0.72 (0.14) & 0.82 (0.14) & 1.00 (0.01) & 1.00 (0.01) & 1.00 (0.01) & 0.99 (0.01) & 0.91 (0.05) \\ 
  100 &  50 &   2 & Unbalanced & 0.83 (0.11) & 0.83 (0.11) & 0.74 (0.16) & 0.73 (0.14) & 0.82 (0.13) & 0.98 (0.02) & 0.98 (0.02) & 0.98 (0.02) & 0.99 (0.01) & 0.91 (0.05) \\ 
  100 &  50 &   3 & Balanced & 0.82 (0.12) & 0.82 (0.11) & 0.66 (0.16) & 0.73 (0.14) & 0.83 (0.13) & 1.00 (0.01) & 1.00 (0.01) & 1.00 (0.01) & 0.99 (0.01) & 0.91 (0.06) \\ 
  100 &  50 &   3 & Unbalanced & 0.82 (0.12) & 0.82 (0.12) & 0.68 (0.16) & 0.72 (0.14) & 0.82 (0.14) & 0.99 (0.02) & 0.99 (0.01) & 0.99 (0.01) & 0.99 (0.01) & 0.91 (0.05) \\ 
  100 & 100 &   2 & Balanced & 0.80 (0.12) & 0.80 (0.11) & 0.60 (0.17) & 0.73 (0.14) & 0.81 (0.14) & 1.00 (0.01) & 1.00 (0.00) & 1.00 (0.00) & 0.99 (0.01) & 0.93 (0.05) \\ 
  100 & 100 &   2 & Unbalanced & 0.80 (0.12) & 0.80 (0.11) & 0.71 (0.16) & 0.72 (0.13) & 0.81 (0.14) & 0.98 (0.03) & 0.98 (0.02) & 0.98 (0.02) & 0.99 (0.01) & 0.92 (0.05) \\ 
  100 & 100 &   3 & Balanced & 0.79 (0.12) & 0.80 (0.11) & 0.60 (0.17) & 0.73 (0.13) & 0.81 (0.14) & 1.00 (0.01) & 1.00 (0.00) & 1.00 (0.00) & 0.99 (0.01) & 0.93 (0.05) \\ 
  100 & 100 &   3 & Unbalanced & 0.80 (0.12) & 0.81 (0.11) & 0.63 (0.16) & 0.73 (0.14) & 0.81 (0.14) & 0.99 (0.01) & 0.99 (0.01) & 0.99 (0.01) & 0.99 (0.02) & 0.93 (0.05) \\ 
  \bottomrule
  \end{tabular}
  \begin{tablenotes}
    \item Note. (S) indicates subgrouping multi-VAR and (CS) indicates multi-VAR with confirmatory subgrouping. Values in parentheses represent Monte Carlo errors.
  \end{tablenotes}
  \end{threeparttable}
}
\end{table}

\begin{table}[ht]
\centering
\caption{MCC and ARI} 
\label{tab:mcc_ari}
\scalebox{0.5}{
\begin{threeparttable}
\begin{tabular}{rrrrcccccccccc}
  \toprule
  &&&& \multicolumn{10}{c}{Measure}\\
 \cmidrule(lr){5-14} &&&& \multicolumn{5}{c}{MCC} &
    \multicolumn{5}{c}{ARI} \\
 \cmidrule(lr){5-9} \cmidrule(lr){10-14} &&&& \multicolumn{5}{c}{Model} & 
    \multicolumn{5}{c}{Model}\\
 \cmidrule(lr){5-9} \cmidrule(lr){10-14} \multicolumn{1}{c}{Time} &
    \multicolumn{1}{c}{Subjects} &
    \multicolumn{1}{c}{Subgroups} &
    \multicolumn{1}{c}{Balance} &
    \multicolumn{1}{c}{multi-VAR (S)} &  
    \multicolumn{1}{c}{multi-VAR (CS)} &
    \multicolumn{1}{c}{multi-VAR} &
    \multicolumn{1}{c}{S-GIMME} &
    \multicolumn{1}{c}{scGVAR} &
    \multicolumn{1}{c}{multi-VAR (S)} &  
    \multicolumn{1}{c}{multi-VAR (CS)} &
    \multicolumn{1}{c}{multi-VAR} &
    \multicolumn{1}{c}{S-GIMME} &
    \multicolumn{1}{c}{scGVAR}\\
 \midrule
50 &  50 &   2 & Balanced & 0.81 (0.12) & 0.82 (0.09) & 0.68 (0.14) & 0.66 (0.14) & 0.56 (0.12) & 0.94 (0.06) & --- & --- & 0.05 (0.06) & 0.91 (0.08) \\ 
  50 &  50 &   2 & Unbalanced & 0.79 (0.12) & 0.79 (0.11) & 0.69 (0.15) & 0.66 (0.14) & 0.55 (0.12) & 0.94 (0.09) & --- & --- & 0.01 (0.04) & 0.90 (0.09) \\ 
  50 &  50 &   3 & Balanced & 0.80 (0.13) & 0.81 (0.09) & 0.68 (0.14) & 0.66 (0.14) & 0.56 (0.12) & 0.94 (0.05) & --- & --- & 0.02 (0.04) & 0.73 (0.20) \\ 
  50 &  50 &   3 & Unbalanced & 0.78 (0.14) & 0.81 (0.10) & 0.68 (0.14) & 0.66 (0.14) & 0.56 (0.12) & 0.92 (0.08) & --- & --- & 0.02 (0.07) & 0.76 (0.12) \\ 
  50 & 100 &   2 & Balanced & 0.82 (0.14) & 0.84 (0.08) & 0.67 (0.16) & 0.66 (0.14) & 0.58 (0.12) & 0.90 (0.05) & --- & --- & 0.07 (0.08) & 0.96 (0.04) \\ 
  50 & 100 &   2 & Unbalanced & 0.79 (0.14) & 0.81 (0.10) & 0.67 (0.17) & 0.66 (0.14) & 0.58 (0.12) & 0.88 (0.06) & --- & --- & 0.04 (0.07) & 0.94 (0.05) \\ 
  50 & 100 &   3 & Balanced & 0.81 (0.14) & 0.83 (0.09) & 0.67 (0.16) & 0.66 (0.14) & 0.58 (0.12) & 0.92 (0.04) & --- & --- & 0.04 (0.04) & 0.90 (0.10) \\ 
  50 & 100 &   3 & Unbalanced & 0.81 (0.15) & 0.83 (0.08) & 0.66 (0.17) & 0.66 (0.14) & 0.59 (0.12) & 0.90 (0.05) & --- & --- & 0.03 (0.05) & 0.80 (0.09) \\ 
  100 &  50 &   2 & Balanced & 0.88 (0.09) & 0.88 (0.08) & 0.77 (0.11) & 0.80 (0.10) & 0.64 (0.11) & 0.98 (0.04) & --- & --- & 0.21 (0.06) & 0.96 (0.07) \\ 
  100 &  50 &   2 & Unbalanced & 0.84 (0.11) & 0.84 (0.11) & 0.78 (0.16) & 0.80 (0.10) & 0.63 (0.11) & 0.94 (0.10) & --- & --- & 0.14 (0.05) & 0.96 (0.07) \\ 
  100 &  50 &   3 & Balanced & 0.88 (0.09) & 0.88 (0.07) & 0.78 (0.11) & 0.81 (0.10) & 0.63 (0.11) & 0.98 (0.03) & --- & --- & 0.17 (0.05) & 0.89 (0.15) \\ 
  100 &  50 &   3 & Unbalanced & 0.84 (0.10) & 0.85 (0.09) & 0.76 (0.12) & 0.80 (0.11) & 0.64 (0.11) & 0.97 (0.05) & --- & --- & 0.20 (0.09) & 0.86 (0.11) \\ 
  100 & 100 &   2 & Balanced & 0.88 (0.09) & 0.88 (0.07) & 0.75 (0.12) & 0.80 (0.10) & 0.66 (0.11) & 0.97 (0.04) & --- & --- & 0.64 (0.06) & 0.98 (0.03) \\ 
  100 & 100 &   2 & Unbalanced & 0.82 (0.12) & 0.82 (0.11) & 0.76 (0.18) & 0.80 (0.10) & 0.65 (0.11) & 0.90 (0.12) & --- & --- & 0.49 (0.08) & 0.98 (0.02) \\ 
  100 & 100 &   3 & Balanced & 0.87 (0.09) & 0.88 (0.07) & 0.74 (0.12) & 0.80 (0.10) & 0.66 (0.11) & 0.97 (0.03) & --- & --- & 0.49 (0.09) & 0.97 (0.06) \\ 
  100 & 100 &   3 & Unbalanced & 0.86 (0.10) & 0.86 (0.08) & 0.75 (0.13) & 0.80 (0.11) & 0.67 (0.11) & 0.97 (0.04) & --- & --- & 0.45 (0.09) & 0.86 (0.08) \\ 
  \bottomrule
  \end{tabular}
  \begin{tablenotes}
    \item Note. (S) indicates subgrouping multi-VAR and (CS) indicates multi-VAR with confirmatory subgrouping. ARI values for mutli-VAR with confirmatory subgrouping and standard multi-VAR are not displayed because they are one and zero, respectively, across all conditions. Values in parentheses represent Monte Carlo errors.
  \end{tablenotes}
  \end{threeparttable}
}
\end{table}

\begin{table}[ht]
\centering
\caption{Absolute Bias and RMSE} 
\label{tab:bias_rmse}
\scalebox{0.48}{
\begin{threeparttable}
\begin{tabular}{rrrrcccccccccc}
  \toprule
  &&&& \multicolumn{10}{c}{Measure}\\
 \cmidrule(lr){5-14} &&&& \multicolumn{5}{c}{Absoulte Bias} &
    \multicolumn{5}{c}{RMSE} \\
 \cmidrule(lr){5-9} \cmidrule(lr){10-14} &&&& \multicolumn{5}{c}{Model} & 
    \multicolumn{5}{c}{Model}\\
 \cmidrule(lr){5-9} \cmidrule(lr){10-14} \multicolumn{1}{c}{Time} &
    \multicolumn{1}{c}{Subjects} &
    \multicolumn{1}{c}{Subgroups} &
    \multicolumn{1}{c}{Balance} &
    \multicolumn{1}{c}{multi-VAR (S)} &  
    \multicolumn{1}{c}{multi-VAR (CS)} &
    \multicolumn{1}{c}{multi-VAR} &
    \multicolumn{1}{c}{S-GIMME} &
    \multicolumn{1}{c}{scGVAR} &
    \multicolumn{1}{c}{multi-VAR (S)} &  
    \multicolumn{1}{c}{multi-VAR (CS)} &
    \multicolumn{1}{c}{multi-VAR} &
    \multicolumn{1}{c}{S-GIMME} &
    \multicolumn{1}{c}{scGVAR}\\
 \midrule
50 &   50 &    2 & Balanced & 0.021 (0.007) & 0.021 (0.007) & 0.026 (0.007) & 0.017 (0.008) & 0.037 (0.008) & 0.096 (0.021) & 0.096 (0.021) & 0.101 (0.023) & 0.100 (0.018) & 0.112 (0.017) \\ 
  50 &   50 &    2 & Unbalanced & 0.021 (0.007) & 0.021 (0.007) & 0.025 (0.007) & 0.017 (0.008) & 0.037 (0.008) & 0.095 (0.021) & 0.095 (0.021) & 0.101 (0.023) & 0.100 (0.018) & 0.112 (0.017) \\ 
  50 &   50 &    3 & Balanced & 0.021 (0.007) & 0.021 (0.007) & 0.026 (0.007) & 0.018 (0.009) & 0.037 (0.008) & 0.095 (0.022) & 0.094 (0.021) & 0.100 (0.023) & 0.100 (0.018) & 0.112 (0.017) \\ 
  50 &   50 &    3 & Unbalanced & 0.021 (0.007) & 0.020 (0.007) & 0.026 (0.007) & 0.017 (0.008) & 0.037 (0.008) & 0.095 (0.021) & 0.094 (0.020) & 0.099 (0.023) & 0.099 (0.018) & 0.111 (0.017) \\ 
  50 &  100 &    2 & Balanced & 0.023 (0.007) & 0.023 (0.007) & 0.032 (0.007) & 0.017 (0.008) & 0.036 (0.008) & 0.107 (0.020) & 0.106 (0.019) & 0.116 (0.022) & 0.098 (0.018) & 0.108 (0.017) \\ 
  50 &  100 &    2 & Unbalanced & 0.023 (0.007) & 0.023 (0.007) & 0.030 (0.007) & 0.017 (0.008) & 0.037 (0.008) & 0.109 (0.020) & 0.109 (0.020) & 0.117 (0.022) & 0.098 (0.018) & 0.109 (0.017) \\ 
  50 &  100 &    3 & Balanced & 0.023 (0.007) & 0.023 (0.007) & 0.032 (0.007) & 0.016 (0.008) & 0.036 (0.008) & 0.107 (0.020) & 0.106 (0.020) & 0.117 (0.022) & 0.098 (0.018) & 0.108 (0.017) \\ 
  50 &  100 &    3 & Unbalanced & 0.023 (0.007) & 0.023 (0.007) & 0.032 (0.007) & 0.017 (0.008) & 0.037 (0.008) & 0.106 (0.020) & 0.106 (0.019) & 0.115 (0.022) & 0.099 (0.018) & 0.109 (0.016) \\ 
  100 &   50 &    2 & Balanced & 0.018 (0.007) & 0.018 (0.007) & 0.024 (0.006) & 0.015 (0.006) & 0.031 (0.007) & 0.089 (0.020) & 0.089 (0.020) & 0.093 (0.022) & 0.075 (0.016) & 0.093 (0.015) \\ 
  100 &   50 &    2 & Unbalanced & 0.018 (0.007) & 0.018 (0.007) & 0.021 (0.007) & 0.015 (0.006) & 0.030 (0.007) & 0.089 (0.020) & 0.089 (0.020) & 0.093 (0.022) & 0.074 (0.016) & 0.092 (0.015) \\ 
  100 &   50 &    3 & Balanced & 0.018 (0.007) & 0.018 (0.007) & 0.023 (0.006) & 0.015 (0.006) & 0.030 (0.007) & 0.088 (0.021) & 0.088 (0.021) & 0.092 (0.023) & 0.074 (0.017) & 0.092 (0.015) \\ 
  100 &   50 &    3 & Unbalanced & 0.018 (0.007) & 0.018 (0.007) & 0.022 (0.006) & 0.015 (0.006) & 0.030 (0.006) & 0.087 (0.020) & 0.087 (0.020) & 0.091 (0.021) & 0.074 (0.015) & 0.091 (0.015) \\ 
  100 &  100 &    2 & Balanced & 0.020 (0.007) & 0.020 (0.007) & 0.029 (0.006) & 0.014 (0.006) & 0.030 (0.006) & 0.102 (0.019) & 0.102 (0.019) & 0.109 (0.021) & 0.071 (0.014) & 0.089 (0.014) \\ 
  100 &  100 &    2 & Unbalanced & 0.021 (0.007) & 0.021 (0.007) & 0.025 (0.007) & 0.014 (0.006) & 0.030 (0.006) & 0.104 (0.019) & 0.105 (0.019) & 0.110 (0.022) & 0.072 (0.015) & 0.090 (0.014) \\ 
  100 &  100 &    3 & Balanced & 0.020 (0.007) & 0.020 (0.007) & 0.029 (0.006) & 0.014 (0.006) & 0.030 (0.006) & 0.101 (0.019) & 0.101 (0.019) & 0.110 (0.021) & 0.071 (0.015) & 0.089 (0.014) \\ 
  100 &  100 &    3 & Unbalanced & 0.020 (0.007) & 0.020 (0.007) & 0.028 (0.006) & 0.014 (0.006) & 0.030 (0.006) & 0.101 (0.019) & 0.100 (0.018) & 0.107 (0.021) & 0.071 (0.014) & 0.089 (0.014) \\ 
  \bottomrule
  \end{tabular}
  \begin{tablenotes}
    \item Note. (S) indicates subgrouping multi-VAR and (CS) indicates multi-VAR with confirmatory subgrouping. Values in parentheses represent Monte Carlo errors.
  \end{tablenotes}
  \end{threeparttable}
}
\end{table}

\printbibliography

\end{document}